\documentclass[]{aa}
\usepackage{graphicx,natbib}
\usepackage{amssymb}

\usepackage{psfrag}

\usepackage[dvips]{color}
\usepackage{tikz}

\DeclareGraphicsExtensions{.eps,.ps}

\begin{document}

\title{New detections of embedded clusters in the Galactic halo}

\author{D. Camargo\inst{1}  \and E. Bica\inst{1}  \and C. Bonatto\inst{1}}

\institute{Universidade Federal do Rio Grande do Sul, Departamento de Astronomia, CP\,15051, RS, Porto Alegre 91501-970, Brazil\\
\email{denilso.camargo@ufrgs.br, bica@if.ufrgs.br, charles@if.ufrgs.br}
\mail{denilso.camargo@ufrgs.br}}

\date{Received --; accepted --}

\abstract
{Until recently it was thought that high Galactic latitude clouds were a non-star-forming ensemble. However, in a previous study we reported the discovery of two embedded clusters (ECs) far away from the Galactic plane ($\sim5$ kpc). In our recent star cluster catalogue we provided additional high and intermediate latitude cluster candidates. }
{This work aims to clarify if our previous  detection of star clusters far away from the disc represents just an episodic event or if the star cluster formation is currently a systematic phenomenon in the Galactic halo. We analyse the nature of four clusters found in our recent catalogue and report the discovery of three new ECs with unusually high latitude and distance from the Galactic disc midplane.
}
{The analysis is based on 2MASS and WISE colour-magnitude diagrams (CMDs), and stellar radial density profiles (RDPs). The CMDs are built by applying a field-star decontamination procedure, which uncovers the cluster's intrinsic CMD morphology.}
{ All of these clusters are younger than 5 Myr. The high-latitude ECs C 932, C 934, and C 939 appear to be related to a cloud complex about 5 kpc below the Galactic disc, under the Local arm. The other clusters are above the disc, C 1074 and C 1100 with a vertical distance of $\sim3$ kpc, C 1099 with $\sim2$ kpc, and C 1101 with $\sim1.8$ kpc.}
{ According to the derived parameters there occur ECs located  below and above the disc, which is an evidence of widespread star cluster formation throughout the Galactic halo. Thus, this study represents a paradigm shift, in the sense that a sterile halo becomes now a  host of ongoing star formation. The origin and fate of these ECs remain open. There are two possibilities for their origin, Galactic fountain or infall. The discovery of ECs far from the disc suggests that the Galactic halo is more actively forming stars than previously thought and since most ECs do not survive the \textit{infant mortality} it may be raining stars from the halo into the disc, and/or the halo harbours generations
of stars formed in clusters like those hereby detected.}

\keywords{{\it(Galaxy:)} open clusters and associations: general; {\it Galaxy:} open clusters and associations: individual; {\it Galaxy:} catalogues; {\it ISM:} clouds}

\titlerunning{High-latitude star clusters}

\authorrunning{Camargo, Bica \& Bonatto}

\maketitle

%

\section{Introduction}
\label{sec:1}

Embedded clusters (ECs) are the first evolutionary stage of open clusters and provide a means to explore the stellar content in gas and dust enshrouded complexes \citep[e.g.][]{Tutukov78, Lada03, Camargo11, Camargo12}. In particular distances and ages can be constrained more accurately.

In the Galaxy most young open clusters and embedded clusters are essentially located within the thin disc below $\sim250$ pc from the Galactic plane \citep[e.g.][]{Camargo13, Camargo15c}. However,  we discovered recently two ECs (Camargo 438 and Camargo 439) within a high-latitude cloud \citep[][hereafter Paper I]{Camargo15b} using WISE \citep{Wright10}. These clusters appear to be related to the high latitude cloud HRK 81.4-77.8 \citep{Heiles88}. Subsequently, in an extended version of our cluster catalogue \citep{Camargo16} we found some other ECs projected close to or on high and intermediate Galactic latitude clouds. Such results suggested ongoing star formation in the Galactic halo.

The Galactic halo is populated by HI clouds known as intermediate and high-latitude clouds (HLCs), which are traced by the 21 cm hyperfine transition line \citep{Blitz84, Dickey90, Magnani96}. 
HLCs as a rule present kinematics inconsistent with Galactic rotation, and are designated as high velocity (HVCs) and intermediate velocity clouds (IVCs), in contrast to low velocity ones (LVCs) that populate the Galactic disc \citep{Muller63, Kuntz96, Martin15}. HLCs appear to be common in disc galaxies such as the Milky Way, and some of them show signs of recent or ongoing mergers \citep{Fraternali01, Thilker04, Battaglia06, Heald07, Oosterloo07, Sancisi08, Cresci10}. Regarding their origin, there is no consensus and both Galactic and extragalactic sources have been proposed.

\begin{figure*}
\centering
\begin{minipage}[b]{0.8\linewidth}
\begin{minipage}[b]{0.49\linewidth}
\includegraphics[width=\textwidth]{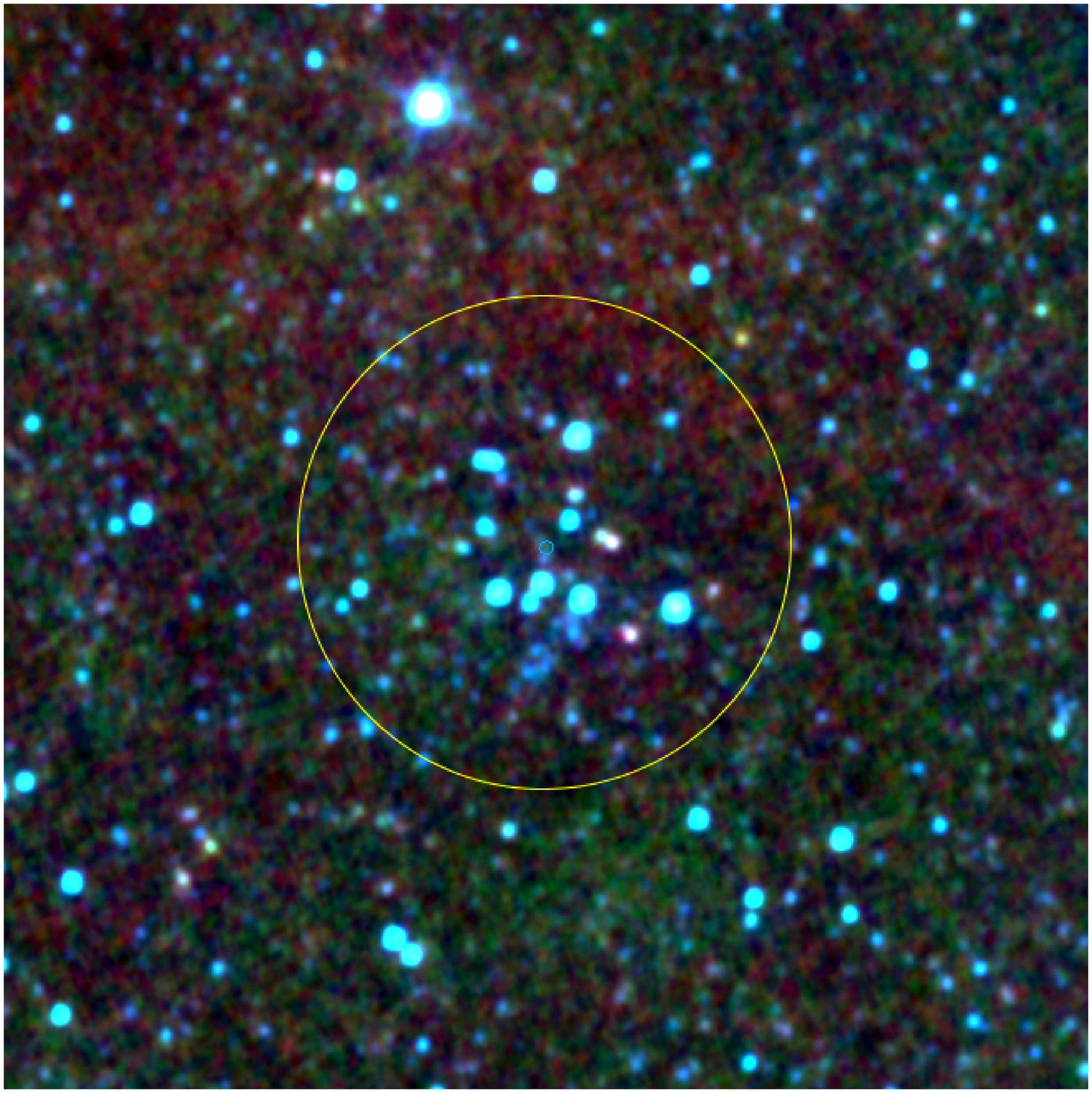}
\put(-180.0,195.0){\makebox(0.0,0.0)[5]{\fontsize{14}{14}\selectfont \color{green} C 1074}}
\end{minipage}\hfill
\vspace{0.02cm}
\begin{minipage}[b]{0.49\linewidth}
\includegraphics[width=\textwidth]{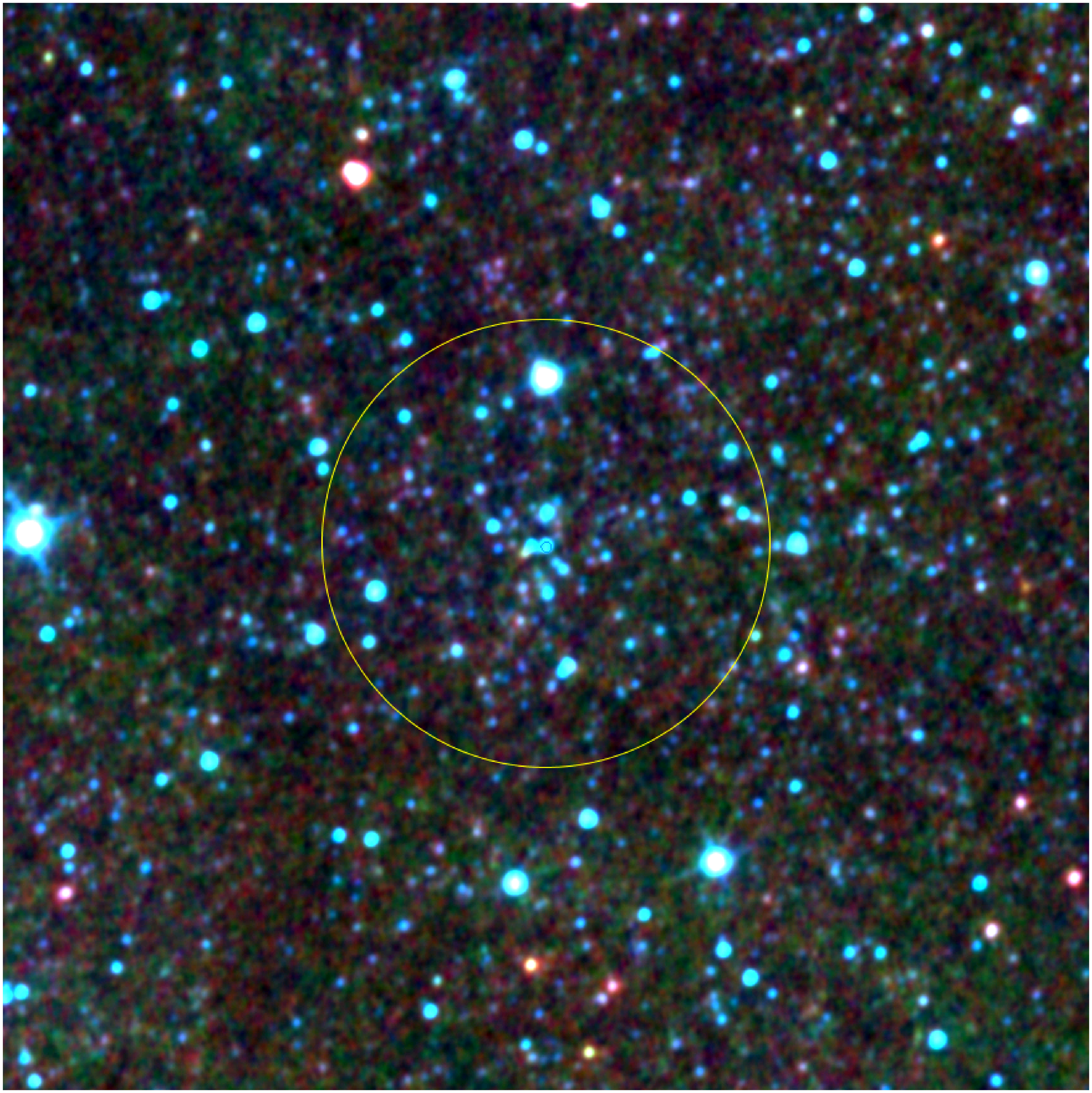}
\put(-180.0,195.0){\makebox(0.0,0.0)[5]{\fontsize{14}{14}\selectfont \color{green} C 939}}
\end{minipage}\hfill
\vspace{0.02cm}
\begin{minipage}[b]{0.49\linewidth}
\includegraphics[width=\textwidth]{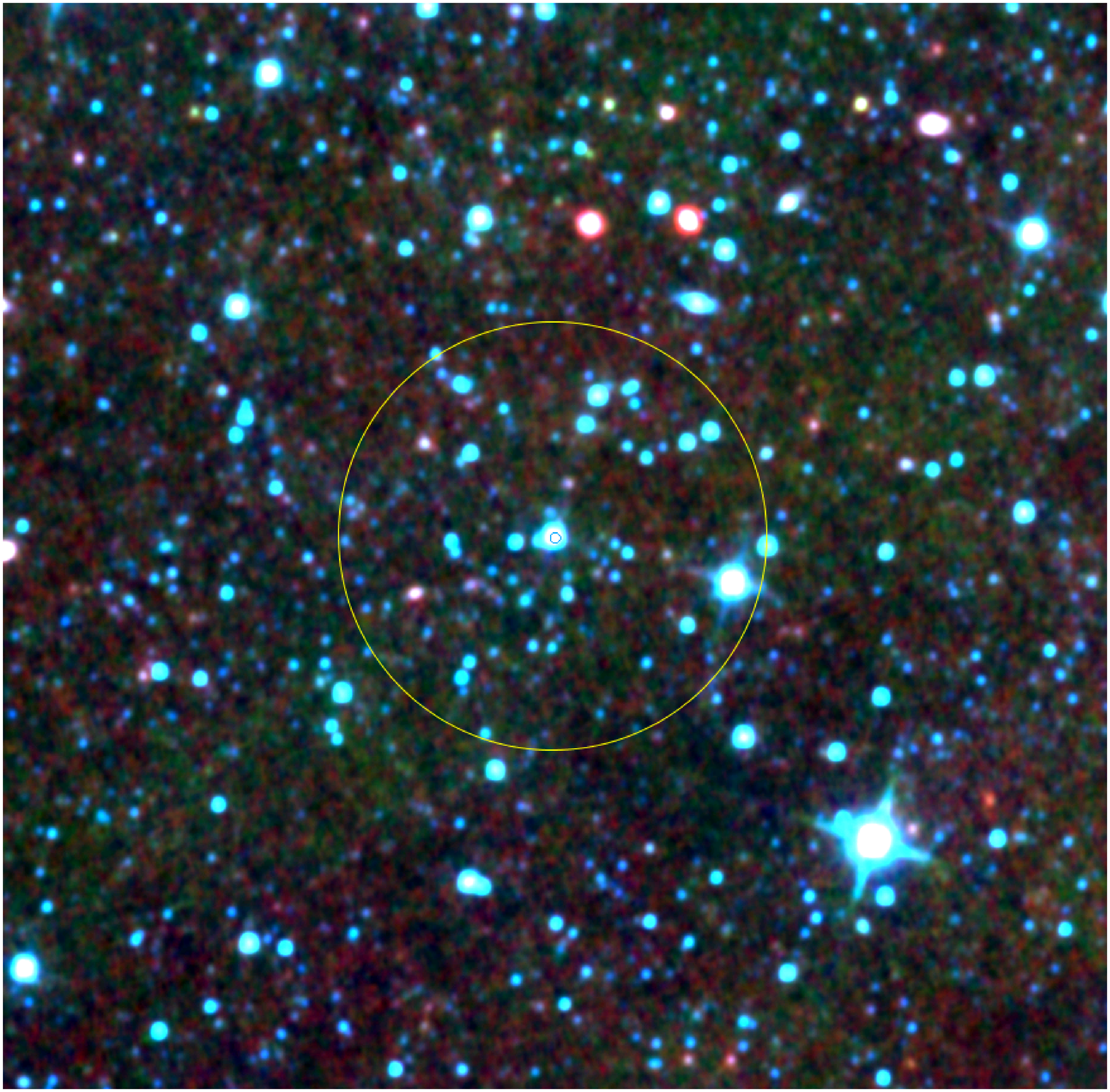}
\put(-180.0,195.0){\makebox(0.0,0.0)[5]{\fontsize{14}{14}\selectfont \color{green}C 1099}}
\end{minipage}\hfill
\vspace{0.02cm}
\begin{minipage}[b]{0.49\linewidth}
\includegraphics[width=\textwidth]{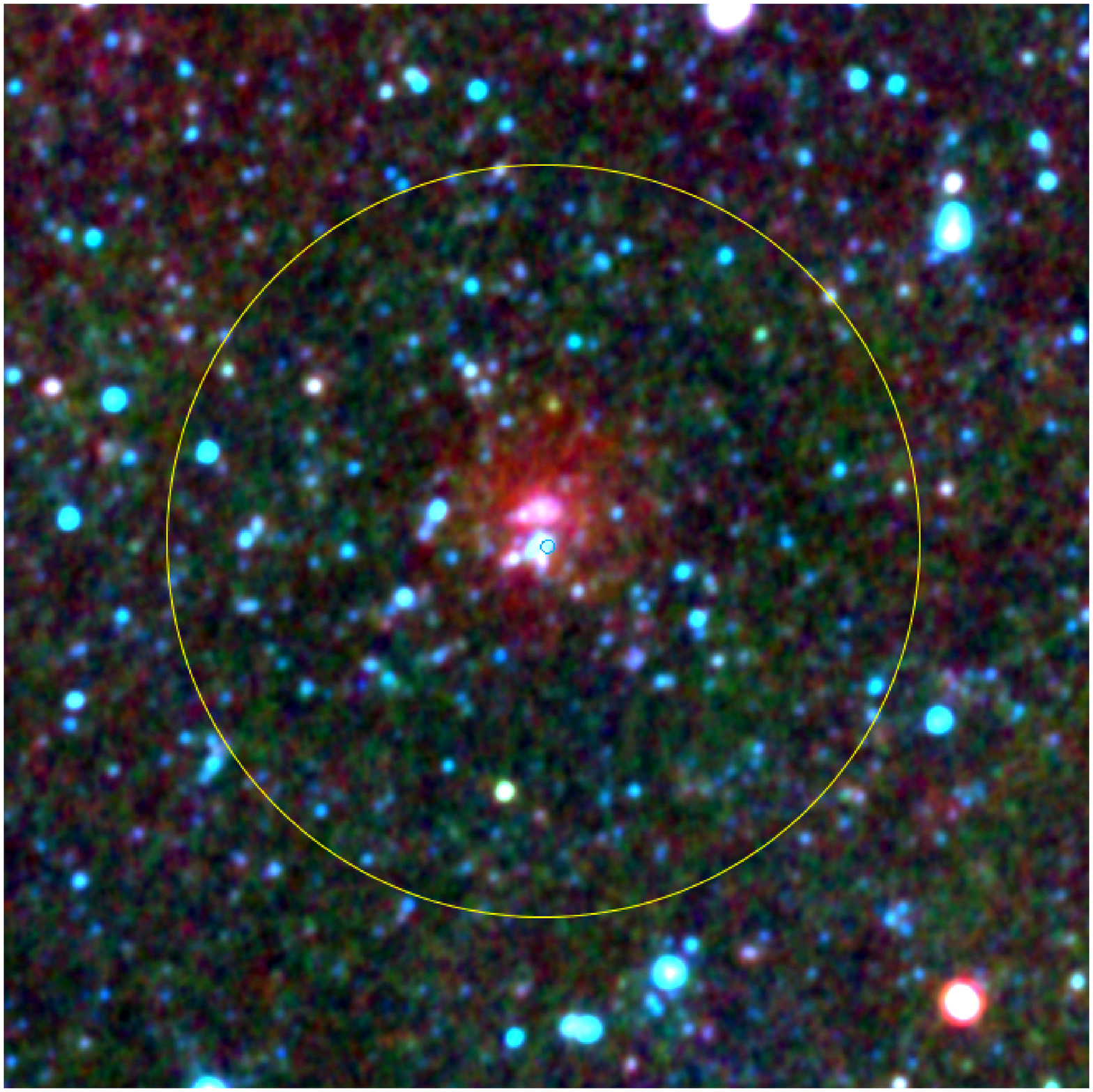}
\put(-180.0,195.0){\makebox(0.0,0.0)[5]{\fontsize{14}{14}\selectfont \color{green}C 934}}
\end{minipage}\hfill
\vspace{0.02cm}
\caption[]{WISE ($15'\times15'$) multicolour images centred on the central coordinates of the embedded clusters C 1074, C 939, C 1099, and C 934. North is to the top and east to the left. Circles encompass more probable cluster stars (Sect. 2).}
\label{f1}
\end{minipage}\hfill
\end{figure*}

\begin{table}[!hb]
\centering
{\footnotesize
\caption{Positions of the present star clusters or candidates.}
\vspace{-0.3cm}
\label{tab1}
\renewcommand{\tabcolsep}{2.6mm}
\renewcommand{\arraystretch}{1,2}
\begin{tabular}{lrrrrr}
\hline
\hline
Target&$\alpha(2000)$&$\delta(2000)$&$\ell$&$b$\\
&(h\,m\,s)&$(^{\circ}\,^{\prime}\,^{\prime\prime})$&$(^{\circ})$&$(^{\circ})$ \\
($1$)&($2$)&($3$)&($4$)&($5$)\\
\hline

C 932 &2:05:02&-18:09:26&188.89&-70.83\\
C 934 &2:05:54&-17:56:17&188.65&-70.54\\
C 939 &2:07:08&-18:13:15&189.83&-70.43\\
C 1074 &10:39:27&-2:00:39&250.15&46.89\\
C 1099 &11:49:55.7&-32:41:42.8&288.23&28.41\\
C 1100 &12:11:39.9&-34:44:45.5&293.73&27.41\\
C 1101 &12:14:24.5&-35:02:04.2&294.41&27.22\\

\hline
\end{tabular}
\begin{list}{Table Notes.}
\item Cols. $2-3$: Central coordinates. Cols. $4-5$: Corresponding Galactic coordinates.
\end{list}
}
\end{table}

HLCs may be produced by an internal engine based on the stellar feedback. In this process winds from OB stars and supernovae blow away gas and dust from the disc in a chimney-like scenario, which subsequently fall back onto the disc as \textit{Galactic fountains} \citep{Shapiro76, Bregman80}. During this phase they can
merge to form molecular clouds. \textit{Chimneys} powered by multiple supernovae within OB associations may blow superbubbles, which can throw gas/dust away on kiloparsec scales \citep{Quilis01, Pidopryhora07}. In this sense, \citet{Melioli08} point out that  typical Galactic OB associations with 100 SNe may eject gas/dust up to $\sim2$ kpc \citep{Quilis01, Pidopryhora07}, but \citet{Melioli09} argue that even multiple OB association cannot throw dust beyond 3.5 kpc.

There are two possibilities for extragalactic origin, \textit{(i)} primordial cold dark-matter encapsulated clouds  accreted directly from the intergalactic medium or \textit{(ii)} infall of dark-matter free clouds remaining from tidal disruption of dwarf galaxies and galaxy collisions \citep{Oort66, Blitz99, Putman04, Keres05, Kaufmann06, Oosterloo07, Sancisi08, Kaufmann10, Hammer15, Wolfe15, Tepper15}. 
Gas accretion is needed to provide the low-metallicity material required by chemical evolution models \citep{Chiappini01}, since in the \textit{Galactic fountain} apparently the gas falls back onto the Galactic disc close to its original locus \citep{Melioli08, Melioli09} and does not affect the metal abundance \citep{Spitoni09}. Besides, the Galaxy on large timescales is apparently forming stars at a constant rate  \citep{Binney00, Fraternali14}, which implies that its gaseous content is being continuously replenished by infall of low metallicity gas \citep{Fraternali12, Joung12}.

\begin{figure*}
\centering
\begin{minipage}[b]{0.8\linewidth}
\begin{minipage}[b]{0.49\linewidth}
\includegraphics[width=\textwidth]{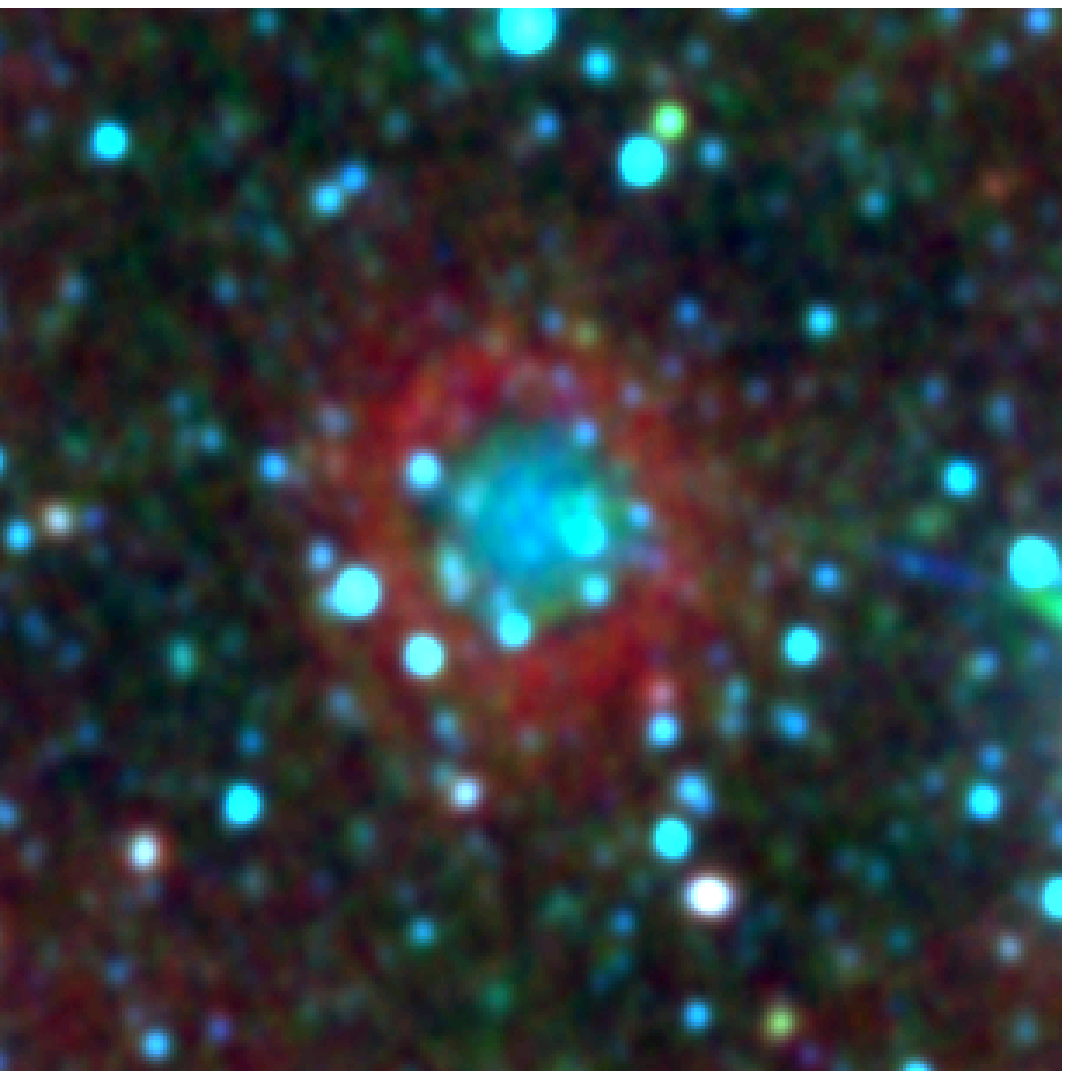}
\end{minipage}\hfill
\begin{minipage}[b]{0.49\linewidth}
\includegraphics[width=\textwidth]{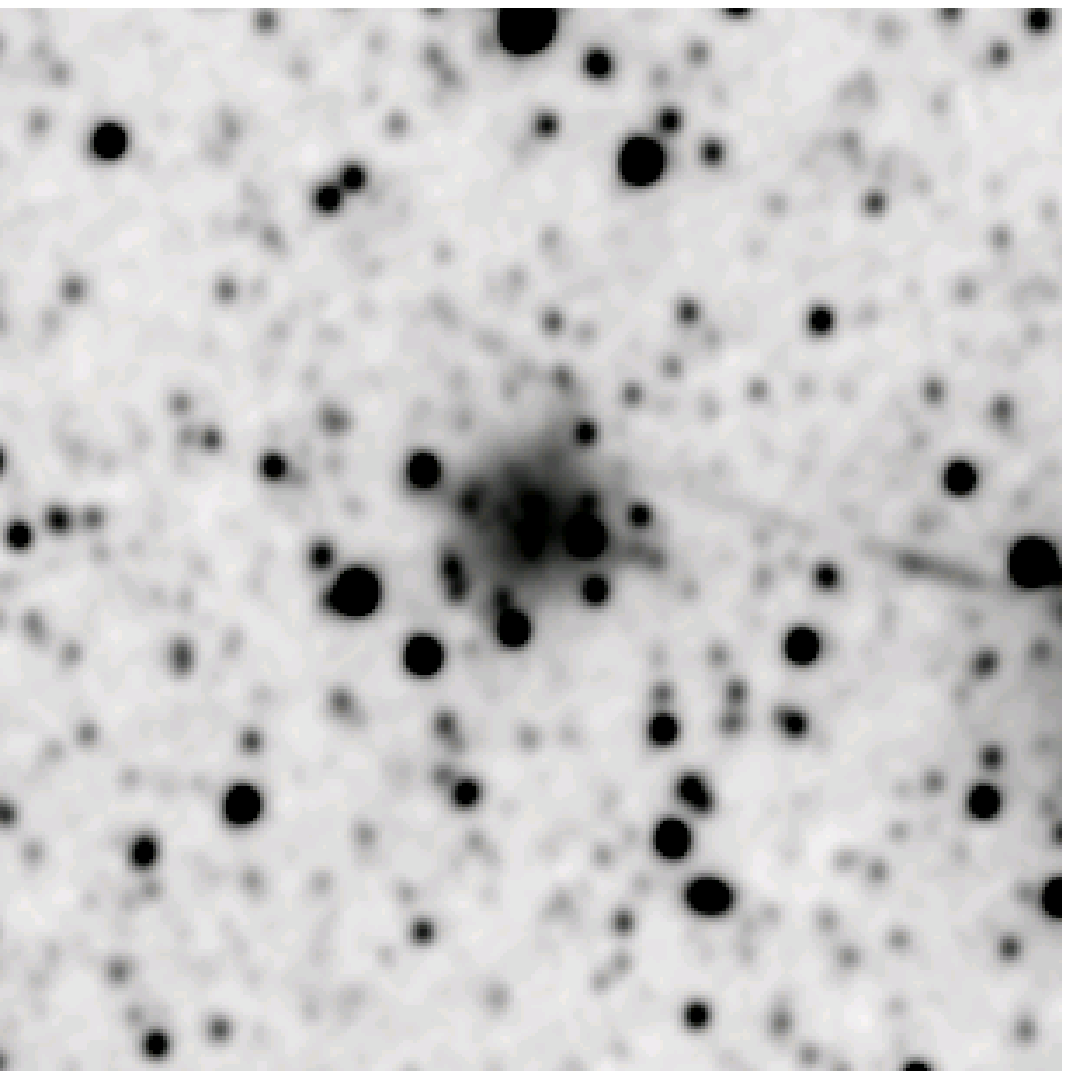}
\end{minipage}\hfill
\caption[]{WISE multicolour and W1 images ($7'\times7'$) centred on C 1100. C 1100 shows a dust emission shell. North is to the top and east to the left.}
\label{f2}
\end{minipage}\hfill
\end{figure*}

\begin{table*}
{\footnotesize
\begin{center}
\caption{Fundamental parameters and Galactocentric components for the ECs in this work.}
\renewcommand{\tabcolsep}{3.4mm}
\renewcommand{\arraystretch}{1.0}
\begin{tabular}{lrrrrrrr}
\hline
\hline
Cluster&$A_V$&Age&$d_{\odot}$&$R_{GC}$&$x_{GC}$&$y_{GC}$&$z_{GC}$\\
&(mag)&(Myr)&(kpc)&(kpc)&(kpc)&(kpc)&(kpc)\\
($1$)&($2$)&($3$)&($4$)&($5$)&($6$)&($7$)&($8$)\\
\hline
C 932 &$1.40\pm0.03$&$2\pm1$&$5.7\pm0.53$&$10.55\pm0.29$&$-9.07\pm0.17$&$-0.29\pm0.03$&$-5.38\pm0.50$\\
C 934 &$1.46\pm0.06$&$2\pm1$&$5.31\pm0.51$&$10.27\pm0.27$&$-8.97\pm0.17$&$-0.27\pm0.03$&$-5.01\pm0.48$\\
C 939 &$1.30\pm0.06$&$3\pm2$&$5.40\pm0.50$&$10.34\pm0.27$&$-9.00\pm0.17$&$-0.31\pm0.03$&$-5.09\pm0.47$\\
C 1074 &$0.93\pm0.06$&$3\pm1$&$4.14\pm0.39$&$9.12\pm0.15$&$-8.18\pm0.09$&$-2.66\pm0.25$&$3.02\pm0.28$\\
C 1099 &$0.71\pm0.06$&$5\pm1$&$4.32\pm0.61$&$7.32\pm0.30$&$-6.03\pm0.17$&$-3.61\pm0.51$&$2.05\pm0.28$\\
C 1100 &$0.93\pm0.06$&$1\pm1$&$6.87\pm0.36$&$8.00\pm0.23$&$-4.76\pm0.13$&$-5.59\pm0.29$&$3.16\pm0.16$\\
C 1101 &$0.96\pm0.06$&$3\pm1$&$3.91\pm0.55$&$6.83\pm0.27$&$-5.78\pm0.20$&$-3.16\pm0.44$&$1.78\pm0.25$\\
\hline
\end{tabular}
\begin{list}{Table Notes.}
\item Col. 2: $A_V$ in the cluster central region. Col. 2: age, from 2MASS photometry. Col. 3: distance from the Sun. Col. 4: $R_{GC}$ calculated using $R_{\odot}=8.3$ kpc as the distance of the Sun to the Galactic centre. Cols. 5 - 8: Galactocentric components. 
\end{list}
\label{tab2}
\end{center}
}
\end{table*}

\begin{table*}
{\footnotesize
\begin{center}
\begin{minipage}[b]{0.7\linewidth}
\caption{Structural parameters for the high latitude embedded cluster C 939.}
\renewcommand{\tabcolsep}{1.9mm}
\renewcommand{\arraystretch}{1.6}
\begin{tabular}{lrrrrrrr}
\hline
\hline
Cluster&$(1')$&$\sigma_{0K}$&$R_{core}$&$R_{RDP}$&$\sigma_{0K}$&$R_{core}$&$R_{RDP}$\\
&($pc$)&($*\,pc^{-2}$)&($pc$)&($pc$)&($*\,\arcmin^{-2}$)&($\arcmin$)&($\arcmin$)\\
($1$)&($2$)&($3$)&($4$)&($5$)&($6$)&($7$)&($8$)\\
\hline
C 939 &$1.36$&$4.1\pm1.2$&$1.8\pm0.4$&$13.6\pm2.0$&$7.65\pm2.15$&$1.32\pm0.29$&$10.0\pm1.5$\\
\hline
\end{tabular}
\begin{list}{Table Notes.}
\item Col. 2: arcmin to parsec scale. To minimise degrees of freedom in RDP fits with the King-like profile, $\sigma_{bg}$ was kept fixed (measured in the respective comparison field) while $\sigma_{0}$ and $R_{core}$ were allowed to vary. 
\end{list}
\label{tab3}
\end{minipage}
\end{center}
}
\end{table*}

There is evidence that IVCs ($|v_{LSR}|=50-100\,km/s$) arise from \textit{Galactic fountains}, while HVCs ($|v_{LSR}|>100\,km/s$) appear to be related to infalling gas \citep{Putman12}. However, if not disrupted on their infall, massive dark-matter free HVCs may be decelerated to LVCs mainly by ram-pressure stripping, Rayleigh-Taylor and Kelvin-Helmholtz instabilities, and dragging forces, leading to substructured clouds \citep{Benjamin97, Maller04, Heitsch09, Plockinger12, Hernandez13}.

 Given the implications of our recent results, a new study was necessary to verify if star formation in the Galactic halo is systematic or an episodic event. Thus, in this work we derive parameters and discuss the properties of seven high and intermediate Galactic latitude ECs.

This paper is organized as follows.
In Sect.~\ref{sect2} we present the cluster sample and derive the fundamental parameters. In Sect.~\ref{sect3} we discuss the results, and in Sect.~\ref{sect4} we provide the concluding remarks.

\begin{figure}
\resizebox{\hsize}{!}{\includegraphics{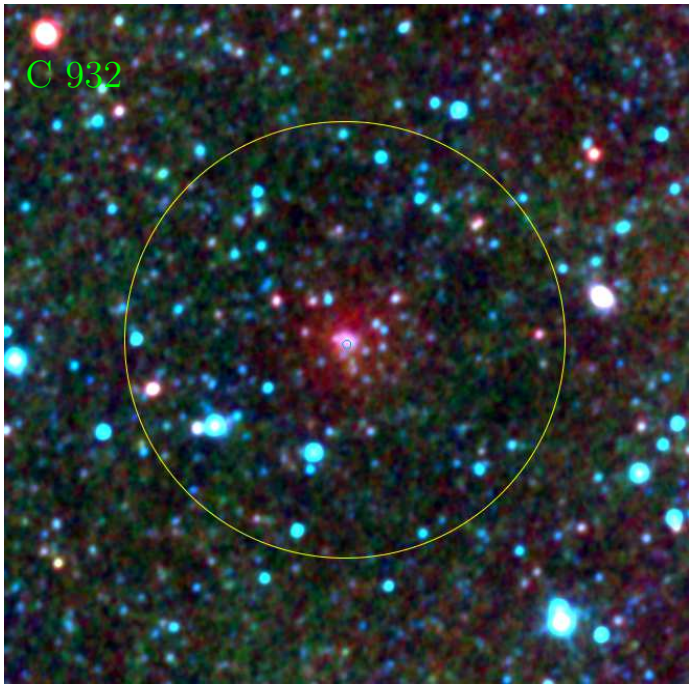}}
\put(-230.0,230.0){\makebox(0.0,0.0)[5]{\fontsize{14}{14}\selectfont \color{green} C 932}}
\caption[]{WISE multicolour image ($15'\times15'$) centred on C 932. C 932 is deeply embedded. North is to the top and east to the left.}
\label{f2b}
\end{figure}

\begin{figure}
\resizebox{\hsize}{!}{\includegraphics{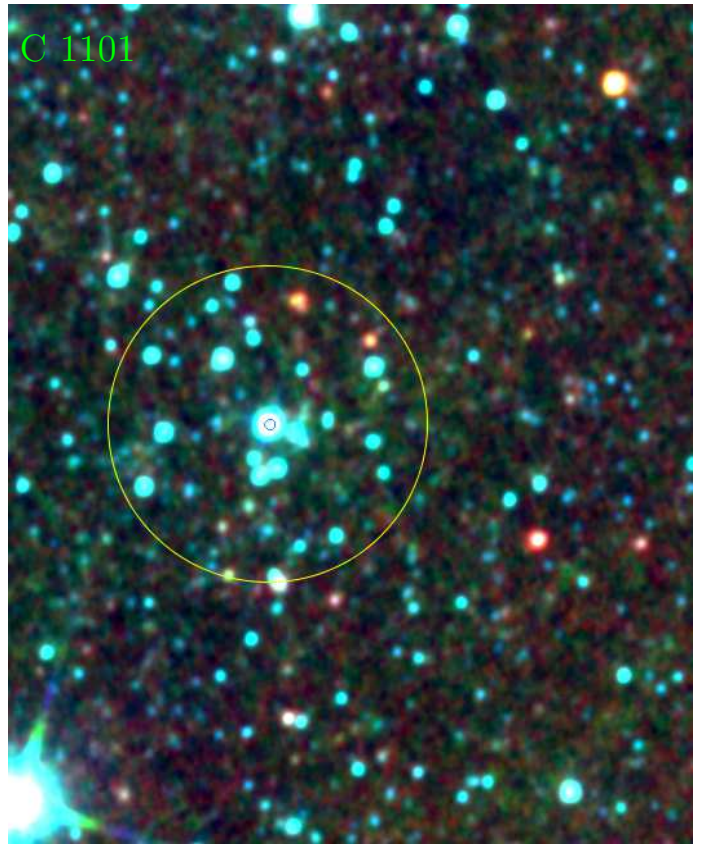}}
\put(-230.0,300.0){\makebox(0.0,0.0)[5]{\fontsize{14}{14}\selectfont \color{green} C 1101}}
\caption[]{WISE ($15'\times15'$) multicolour images centred on the central coordinates of the embedded cluster C 1101. North is to the top and east to the left.}
\label{f3}
\end{figure}

\begin{figure}
\centering
\begin{minipage}[b]{0.85\linewidth}
\resizebox{\hsize}{!}{\includegraphics{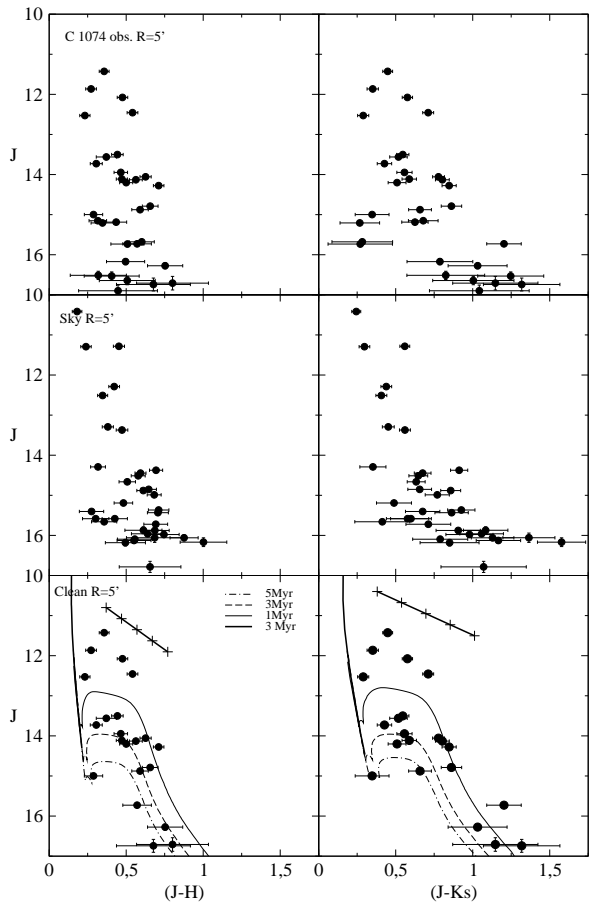}}
\end{minipage}
\vspace{-0.2cm}
\caption[]{C 1074: example of decontamination procedure. Top panels: 2MASS observed CMDs. Middle panels: equal area comparison field. Bottom panels: field-star decontaminated CMDs fitted with PARSEC isochrones for MS and PMS stars. We also show the reddening vector for $A_v=1$ to $5$.}
\label{cmd1a}
\end{figure}

\section{Cluster analysis} 
\label{sect2}

We searched for clusters by looking for stellar density enhancements in the WISE Atlas, initially in large areas ($2^{\circ}\times2^{\circ}$) for the identification of possible candidates. Subsequently, we inspected these candidates in more detail using the WISE individual bands. Heavily dust obscured ECs are more easily detected in the W3 ($12{\mu}m$) and W4 ($22{\mu}m$) WISE bands that are more sensitive to dust emission. W1 ($3.4{\mu}m$) and W2 ($4.6{\mu}m$) are more adequate to identify the stellar content. The multicolour (false colours) images are a composition of the four bands. Given the rarity of the new findings, the search required a long time and dedication. The initial inspection and selection of a sample of cluster candidates were made by one of us (D.C) and subsequently checked independently by the other authors.

\begin{figure}
\centering
\begin{minipage}[b]{0.85\linewidth}
\resizebox{\hsize}{!}{\includegraphics{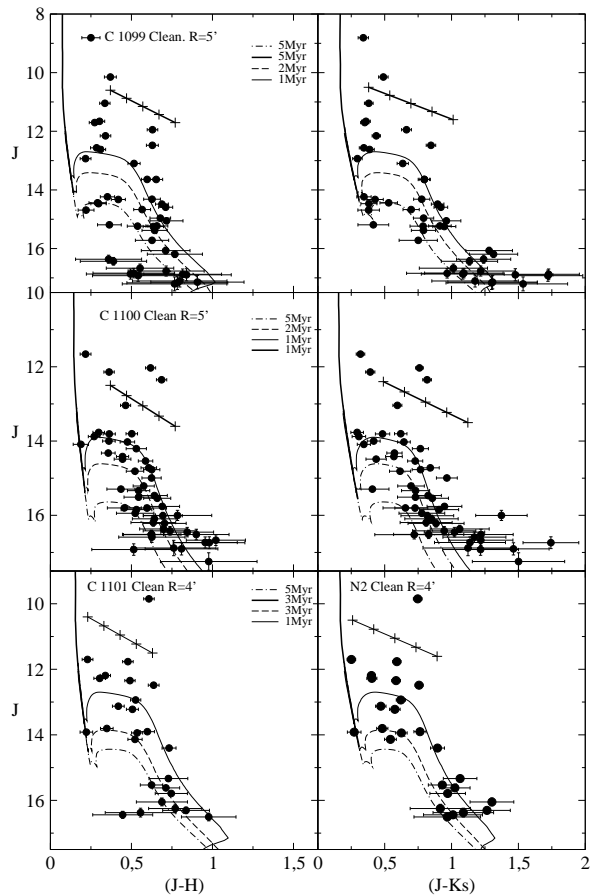}}
\end{minipage}
\vspace{-0.2cm}
\caption[]{Decontaminated CMDs fitted by PARSEC isochrones, for the newly found embedded cluster C 1099 (\textit{top panels}), C 1100 (\textit{middle panels}), and C 1101 (\textit{bottom panels}). We also show the reddening vector for $A_V=1$ to $5$.}
\label{cmd1}
\end{figure}

\begin{figure}
\centering
\begin{minipage}[b]{0.85\linewidth}
\resizebox{\hsize}{!}{\includegraphics{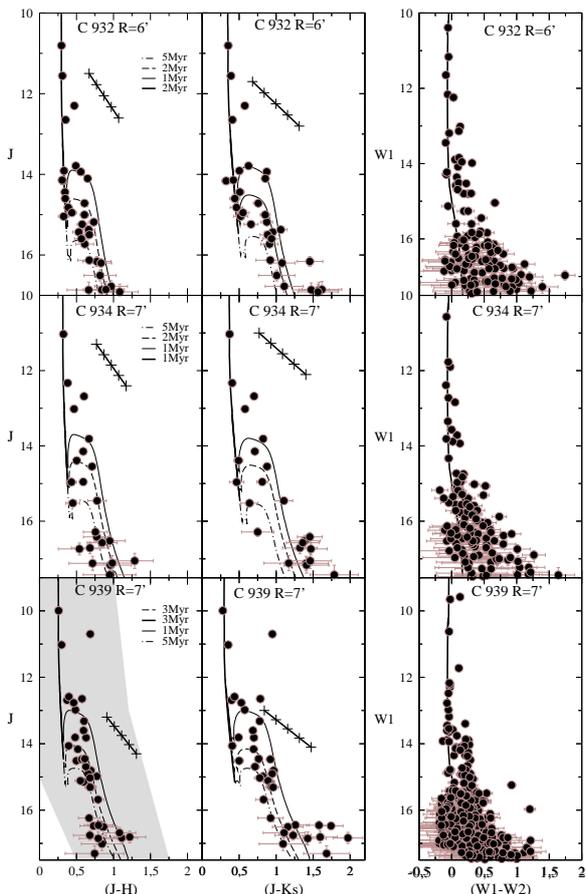}}
\end{minipage}
\vspace{-0.2cm}
\caption[]{Decontaminated CMDs for the high-latitude embedded cluster C 932 (\textit{top panels}), C 934 (\textit{middle panels}), and C 939 (\textit{bottom panels}). The colour-magnitude filter used to isolate cluster
stars is shown as a shaded region in the decontaminated CMD of C 939. We also show the reddening vector for $A_V=1$ to $5$.}
\label{cmd2}
\end{figure}

\begin{figure}
\centering
\begin{minipage}[b]{0.85\linewidth}
\resizebox{\hsize}{!}{\includegraphics{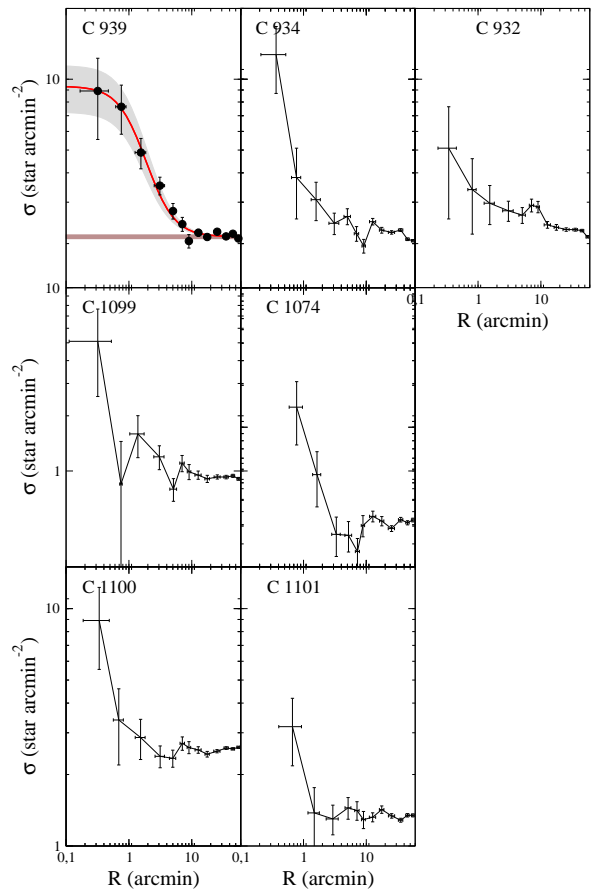}}
\end{minipage}
\vspace{-0.2cm}
\caption[]{Radial density profile for the embedded clusters in this study. The brown horizontal shaded region in the RDP of C 939 corresponds to the stellar background level measured in the comparison field and the grey region correspond to the $1\sigma$ King fit uncertainty.}
\label{rdp}
\end{figure}

\begin{figure}[!hb]
\vspace{-0.3cm}
\resizebox{\hsize}{!}{\includegraphics{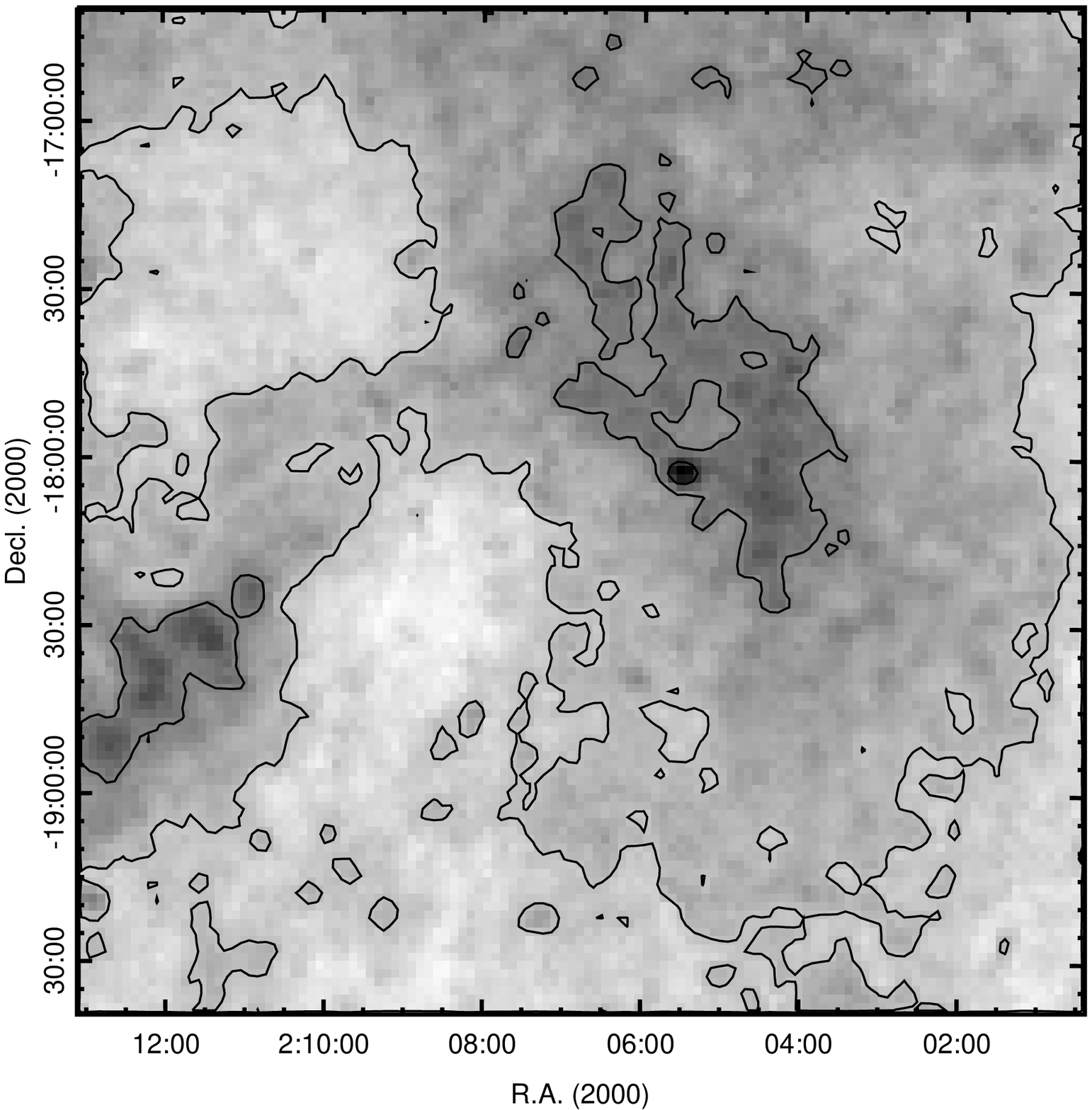}}
\put(-129.0,120.0){\makebox(0.0,0.0)[5]{\fontsize{15}{15}\selectfont \color{green}$\bullet$}}
\put(-98.2,129.5){\makebox(0.0,0.0)[5]{\fontsize{15}{15}\selectfont \color{green}$\bullet$}}
\put(-112.0,143.0){\makebox(0.0,0.0)[5]{\fontsize{15}{15}\selectfont \color{green}$\bullet$}}
\vspace{-0.2cm}
\caption[]{IRAS $100{\mu}m$ image of the cloud complex possibly related to the high-latitude ECs C 932, C934, and C 939, which are identified by green circles.}
\label{cloud1}
\end{figure}

The sample of star clusters projected towards the halo is listed in Table \ref{tab1}, together with their equatorial and Galactic coordinates.
Three of them are ECs discovered in the present study by means of visual inspection on WISE images and four were previously reported in \citet{Camargo16}. 
Following our recent star cluster catalogue designations \citep{Camargo15a, Camargo15c, Camargo16} we adopt the names C 1099, C 1100 and C 1101 for the newly discovered ECs. Fig.~\ref{f1} shows multicolour images of four sample clusters. They are considerably populated, concentrated, and contrasted. Fig.~\ref{f2} shows images of C 1100 in multicolour and W1, while Fig.~\ref{f2b} gives multicolour image for C 932. C 1100 is surrounded by dust emission, while C 932 is deeply embedded. Fig.~\ref{f3} presents a WISE multicolour image of C 1101. CMDs and RDPs are shown in Figs.~\ref{cmd1a} to \ref{rdp} while Figs.~\ref{cloud1} and \ref{cloud2} give the positions of three clusters (C 932, C 934, and C 939) that are projected on a halo cloud as seen in IRAS $100{\mu}m$ and WISE multicolour, respectively. As far as we are aware the halo cloud containing the 3 new clusters has not been reported before. We will refer to it as CBB 188.13-70.84, following \citet{Heiles88} for the cloud HRK 81.44-77.8, which  was the first halo cloud with detected star formation (Paper I). The stellar density contrast between clusters and field is evident in Fig.~\ref{cloud2}. Note that the diffuse dust emission of the cloud CBB 188.13-70.84 is stronger to the NE corner of Fig.~\ref{cloud2}, in agreement with Fig.~\ref{cloud1}. Finally, Fig.~\ref{Milky1} shows the spatial distribution of the clusters in this study.

To disentangle the cluster intrinsic evolutionary sequences in the CMD from those of the stellar background, we built the CMDs by applying a field star decontamination procedure to the raw photometry extracted from a circular area centred on each cluster \citep{Bonatto07, Bonatto08, Bonatto10, Bica08a}.
Cluster fundamental parameters are derived with PARSEC isochrones \citep{Bressan12} fitted to the cluster sequences in the decontaminated CMDs \citep{Camargo13}. The clusters present a standard MS and PMS distribution in the CMDs \citep[e.g.][]{Bonatto09}.

Fig.~\ref{cmd1a} gives the 2MASS $J\times(J-H)$ and $J\times(J-K_s)$ CMDs for C 1074. We illustrate the background stellar density subtraction. The top panels show the raw photometry extracted from the cluster central region ($R=5$ arcmin). The middle panels show the equal area comparison fields. In the bottom panels we show the decontaminated CMDs and fit them with MS and PMS isochrones. The best fitting isochrones provide an age of $3\pm1$ Myr for a distance from the Sun of $\sim4.1$ kpc and distance from the Galactic disc mid-plane of $\sim3$ kpc.
Fig.~\ref{cmd1} shows the 2MASS CMDs for the newly found ECs C 1099, C 1100, and C 1101. For brevity we show only the $J\times(J-H)$ and $J\times(J-K_s)$ field-star decontaminated CMDs. Table~\ref{tab2} gives ages, distance from the Sun, and vertical distance from the midplane.

We adopt WISE photometry to build radial density profiles (RDPs), since the WISE bands are more sensitive to PMS stars. 
The RDPs of the present objects (Fig.~\ref{rdp}) are typical of clusters in the embedded phase and do not follow a King's law \citep{Bonatto11, Camargo10, Camargo11, Camargo12}, except C 939. For this object we built the RDP with stars selected by a colour-magnitude filter (Fig.~\ref{cmd2}) used to discard stars with colours compatible with those of the field stars \citep{StrucPar, Camargo09}. We tentatively built colour magnitude filters for all objects in this study, but the filtered photometry did not produce significant gains with respect to the raw photometry. Since stars of different masses in the decontaminated CMDs of ECs in this work can be very reddened, such filters are not able to discard a significant amount of field stars without affecting the cluster PMS sequence.
We fitted the RDP of C 939 with a King's function $\sigma(R)=\sigma_{bg}+\sigma_{0}/(1+(R/R_{core})^{2}$ \citep{King62}, where $\sigma_{bg}$ is the residual background surface density of stars, $\sigma_{0}$ is the central density of stars and $R_{core}$ is the core radius. The cluster radius ($R_{RDP}$) and uncertainty can be estimated considering the fluctuations of the RDPs with respect to the residual background. $R_{RDP}$ corresponds to the distance from the cluster centre where the RDP and the comparison field stellar density become statistically indistinguishable. $\sigma_{bg}$ is measured in the comparison field and kept fixed.
The RDP of C 939 gives a cluster radius of $13.6\pm2$ pc and a central density of $\sigma_{0}=4.1\pm1.2^{*\,pc^{-2}}$. It is probable that C 939 simply mimics a King's profile, owing to its young age.
Table~\ref{tab3} provides structural parameters derived for the particular case of C 939. The remaining clusters have profiles that indicate the presence of a cluster, but ECs in the early evolutionary phase as a rule lack the span of time to be an isothermal sphere and thus cannot be fitted by a King-like profile \citep{Camargo12, Camargo15c, Bica08b}. Such profiles fit clusters in near dynamical equilibrium. On the other hand, some ECs appear to be evolved enough to follow or simply to mimic a King law \citep{Camargo10, Camargo11}.

\section{Discussion}
\label{sect3}

The discovery of high latitude ECs is fundamental to our understanding of the Galaxy formation, evolution and dynamics, since the new findings affect the role of the halo in the Galactic evolutionary process. It seems more active than previously thought. If these clusters behave like those formed in the Galactic disc \citep[see \textit{infant mortality -}][]{Tutukov78, Lada03}, most of them will not reach the disc as bound systems. Thus, it may be raining young stars from the halo into the disc. Alternatively, generations of stars, as formed in the present clusters, may be populating the halo.

\begin{figure*}[!ht]
\resizebox{\hsize}{!}{\includegraphics{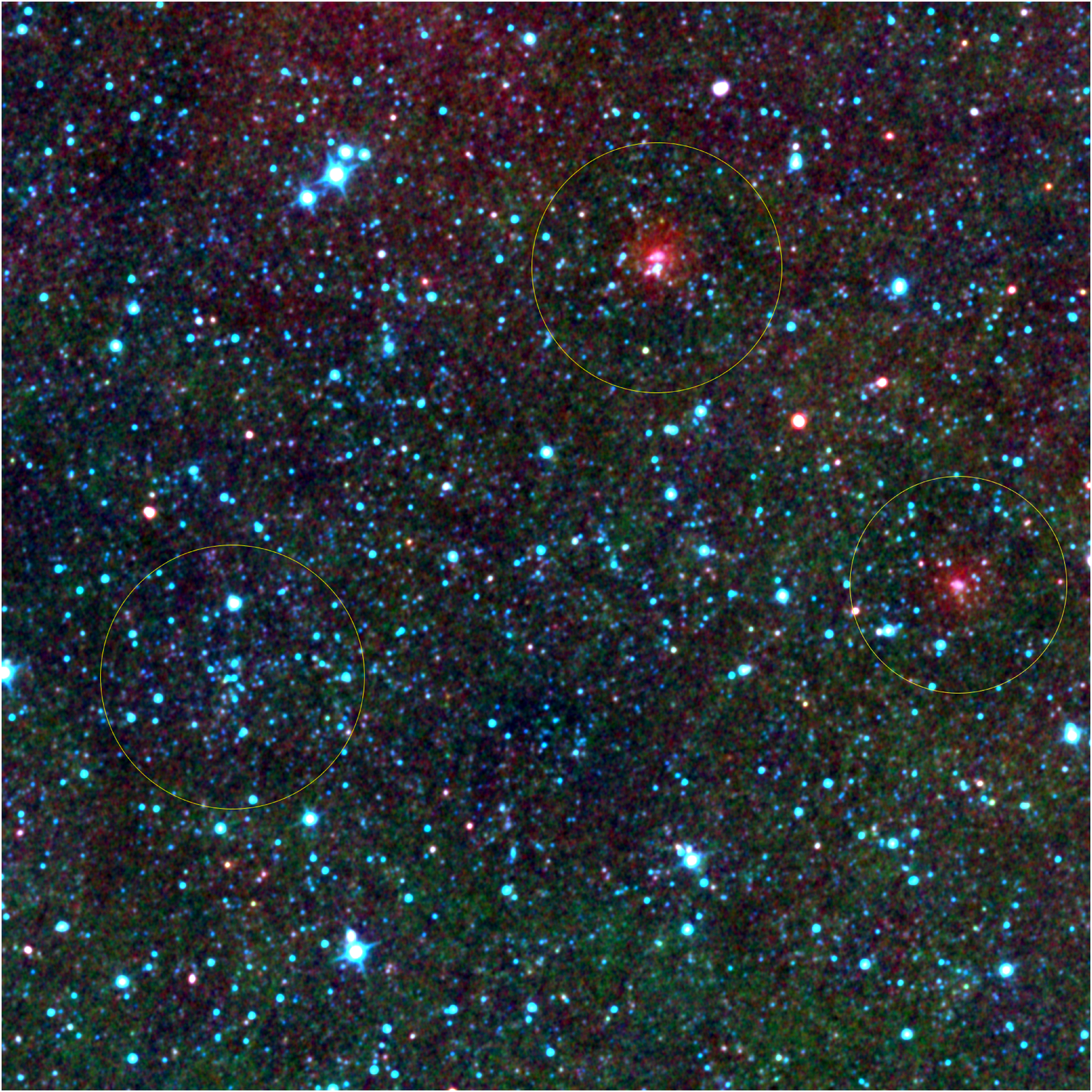}}
\put(-210.0,470.0){\makebox(0.0,0.0)[5]{\fontsize{15}{15}\selectfont \color{green} C 934}}
\put(-60.0,310.0){\makebox(0.0,0.0)[5]{\fontsize{15}{15}\selectfont \color{green} C 932}}
\put(-415.0,275.0){\makebox(0.0,0.0)[5]{\fontsize{15}{15}\selectfont \color{green} C 939}}
\caption[]{WISE multicolour image ($45'\times45'$) with a zoom in on the cloud complex of Fig.~\ref{cloud1} to show the ECs C 932, C934, and C 939 in more detail.}
\label{cloud2}
\end{figure*}

\begin{figure*}[!ht]
\centering
\begin{minipage}[b]{0.98\linewidth}
\begin{minipage}[b]{0.49\linewidth}
\resizebox{\hsize}{!}{\includegraphics{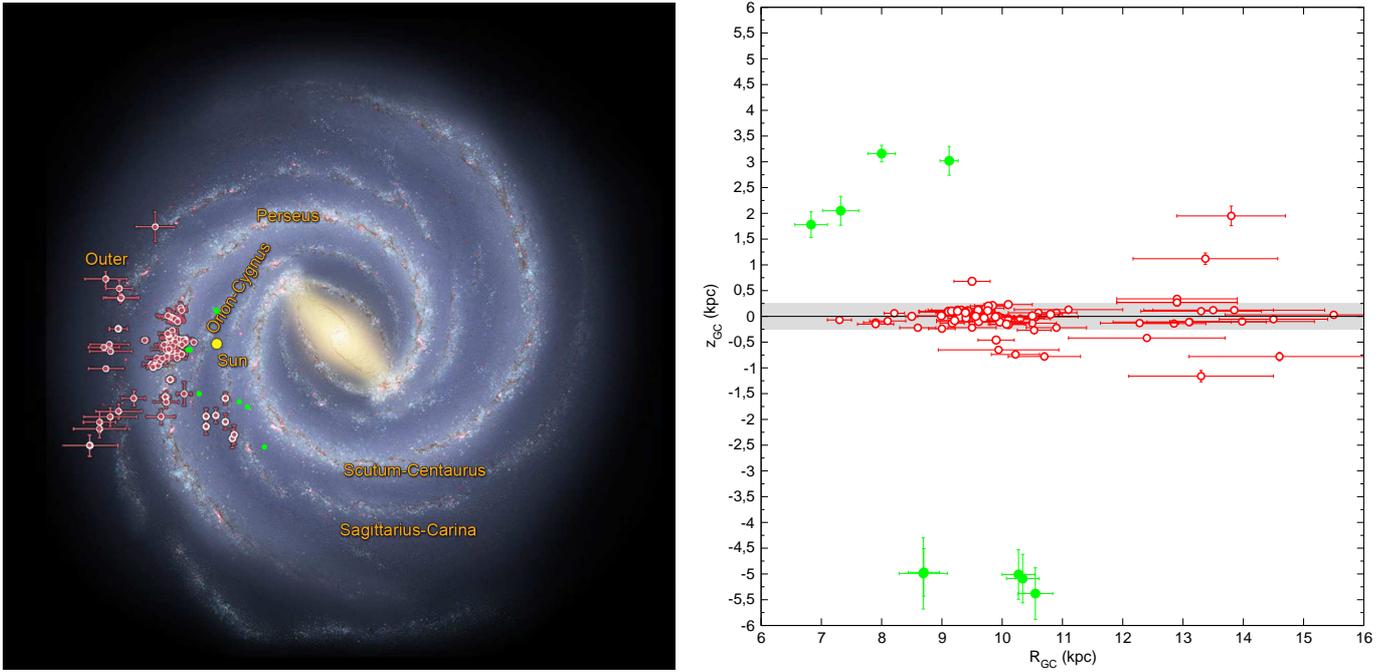}}
\put(-181.4,120.6){\makebox(0.0,0.0)[5]{\fontsize{90}{00}\selectfont \color{green}.}}
\put(-182.2,120.6){\makebox(0.0,0.0)[5]{\fontsize{90}{00}\selectfont \color{green}.}}
\put(-178.0,104.0){\makebox(0.0,0.0)[5]{\fontsize{90}{90}\selectfont \color{green}.}}
\put(-153.5,83.8){\makebox(0.0,0.0)[5]{\fontsize{90}{90}\selectfont \color{green}.}}
\put(-171.4,135.0){\makebox(0.0,0.0)[5]{\fontsize{90}{90}\selectfont \color{green}.}}
\put(-171.6,135.5){\makebox(0.0,0.0)[5]{\fontsize{90}{90}\selectfont \color{green}.}}
\put(-162.9,101.0){\makebox(0.0,0.0)[5]{\fontsize{90}{90}\selectfont \color{green}.}}
\put(-159.9,99.0){\makebox(0.0,0.0)[5]{\fontsize{90}{90}\selectfont \color{green}.}}
\end{minipage}\hfill
\begin{minipage}[b]{0.49\linewidth}
\resizebox{\hsize}{!}{\includegraphics{testeGalaxy.eps}}
\end{minipage}\hfill
\end{minipage}\hfill
\caption[]{Spatial distribution of the ECs in this study and Paper I (green circles) compared to ECs in our previous works (red circles). Credit: Robert Hurt (NASA/JPL) and \citet{Camargo15c}.}
\label{Milky1}
\end{figure*}

As possible origin for the ECs in this sample one can trace them to Galactic fountains or infall. Both scenarios present possible explanations and restrictions.
The accretion of low metallicity gas from the intergalactic medium may occur through filamentary structures with the gas cooling into clouds \citep{Fernandez12}. Subsequently HVCs may be destroyed or fragmented into smaller clouds by drag forces from the halo and phenomena such as Rayleigh-Taylor and Kelvin-Helmholtz instabilities. Star formation can be triggered by these interactions, within clouds that reach enough density (Figs.~\ref{cloud1} and \ref{cloud2}). However, there is evidence that star formation is possible only within dark-matter encapsulated HVCs such as the Smith Cloud  \citep{Smith63, Heitsch09, Nichols09, Joung12}.  \citet{Christodoulou97} argue that without dark matter, HVCs are unable to reach the mass required to collapse. 
On the other hand, C 932, C934, and C939 are located right above the Local spiral arm (Fig.~\ref{Milky1}), which would be consistent with the chimney scenario. \citet{Schlafly15} mapped various bubble-like structures vertically along the range 0.3 to 2.8 kpc, which form the Orion superbubble. The expansion of these substructures powered by massive stellar winds and supernovae are triggering star formation in various shells and rings, inputting energy to the superbubble \citep{Lee09}. The star formation engine in the Galactic fountain may work in a similar way as the infall scenario, through the interaction of a cloud with the surrounding halo environment.
However, high-latitude clusters in this study are located at distances from the disc larger than expected in recent works for a chimney-like event \citep{Melioli09}.
Regardless of the scenario, a possible cloud-cloud interaction may be leading clouds in Fig.~\ref{cloud1} to collapse and triggering not only star formation, but also cluster formation. However, the timescale for cloud-cloud collision in a cloud complex appears to be larger than 1 Gyr \citep{Christodoulou97}.   
 Thus, additional studies are required to check the presence of dark-matter in HVCs, estimate the cloud-cloud interaction timescales, and to provide more insight on the effect of halo environment on the HVCs and chimney-like events. This is beyond the purpose of this study. 

In Paper I we calculated orbits for the 2 ECs therein studied using  UCAC4 proper motions and a Galactic Potential. We could not decide between a Galactic or extragalactic origin for the cloud, given the uncertainties. With the advent of Gaia \citep{Perryman01}, which will provide very accurate positions and velocities for stars in the Galactic halo, we intend to explore in more detail orbits for the ECs from this study and Paper I. The halo has been site in recent years of discoveries of low luminosity dwarf galaxies, tidal streams, and faint globular clusters \citep[e.g.][]{Antoja15, Bechtol15, Drlica15, Huxor15, Koposov15, Luque16}.
Now, the present study and Paper I indicate that yet another process occurs in the halo: recent star formation.

In \citep{Camargo15b}, we discovered star cluster formation in the Galactic halo. Despite the importance of that discovery, there remained some questions, for instance,  would it be just an episodic event of star cluster formation far away from the disc, or instead the halo would be a site of systematic EC formation? Subsequently, in \citet{Camargo16} we found four additional high latitude ECs, and here we analyse them together with three others found in this study.
 Thus, this work points to a new paradigm in the star and star cluster formation, in the sense that the formation of such objects occurs in the halo and it seems to be systematic.

The embedded cluster C 1100 shows a dust emission bubble structure (Fig.~\ref{f2}). Such objects with a bubble are relatively common in our recent catalogue \citep{Camargo16}, especially for ECs and embedded stellar groups with ongoing star formation. See also the EC catalogue by \citet{Majaess13}.

\section{Concluding remarks}
\label{sect4}

In Paper I we reported two ECs (C 438 and C 439) within a high-latitude cloud, which were the first high-latitude embedded clusters discovered. In this study we find new results about star cluster formation in high-latitude clouds and further infer that the Galactic halo is currently forming stars within ECs. 

Using 2MASS and WISE photometry we analysed the nature of seven ECs at high and intermediate Galactic latitudes, three of them first reported in this work (C 1099, C 1100, and C 1101). The age of these clusters are in the range of 1 to 5 Myr. C 932, C 934, and C 939 are high-latitude ECs projected within the newly identified cloud complex including CBB 188.13-70.84. These clusters are located at a vertical distance of about 5 kpc below the Galactic disc. C 1074, C 1099, C 1100, and C 1101 are in the range 1.7 to 3.2 kpc above the disc. The clusters show decontaminated CMDs with the typical pattern of MS and PMS stars in embedded clusters (Paper I and references therein). Their spatial distribution is given in Fig.~\ref{Milky1}. We have gathered a significant collection of very young star clusters in the halo.

Paper I and the present additional study point to a paradigm shift in the halo, which becomes an ongoing site of star formation in the Galaxy. 

\vspace{0.8cm}

\textit{Acknowledgements}: We thank an anonymous referee for valuable comments and suggestions. 
This publication makes use of data products from the Two Micron All Sky Survey (2MASS) and Wide-field Infrared Survey Explorer (WISE). The 2MASS is a joint project of the University of Massachusetts and the Infrared Processing and Analysis Centre/California Institute of Technology, funded by the National Aeronautics and Space Administration and the National Science Foundation. WISE is managed and operated by NASA’s Jet Propulsion Laboratory (JPL) in Pasadena, California and is a project of the JPL/California Institute of Technology, funded by the National Aeronautics and Space Administration. The spacecraft scanned the entire sky twice. E. Bica and C. Bonatto acknowledge support from CNPq (Brazil).


\begin{thebibliography}{}

\bibitem[\protect\citeauthoryear{Antoja et al.}{2015}]{Antoja15} 
    Antoja, T., Mateu, C., Aguilar, L., Figueras, F., Antiche, E., Hern{\'a}ndez-P{\'e}rez, F., Brown, A.~G.~A., Valenzuela, O., et al. 2015, MNRAS, 453, 541

\bibitem[\protect\citeauthoryear{Battaglia et al.}{2006}]{Battaglia06} 
    Battaglia, G., Fraternali, F., Oosterloo, T. \& Sancisi, R. 2006, A\&A, 447, 49

\bibitem[\protect\citeauthoryear{Bechtol et al.}{2015}]{Bechtol15} 
    Bechtol, K., Drlica-Wagner, A., Balbinot, E., Pieres, A., Simon, J.~D., Yanny, B., Santiago, B., Wechsler, R.~H., et al. 2015, ApJ, 807, 50

\bibitem[\protect\citeauthoryear{Benjamin \& Danly}{1997}]{Benjamin97} 
    Benjamin, R.~A. \& Danly, L. 1997, ApJ, 481, 764

\bibitem[\protect\citeauthoryear{Bica et al.}{2008a}]{Bica08a} 
   Bica, E., Bonatto, C. \& Camargo, D.  2008a, MNRAS, 385, 349

\bibitem[\protect\citeauthoryear{Bica et al.}{2008b}]{Bica08b} 
   Bica, E., Bonatto, C. \& Dutra, C. M.  2008b, A\&A, 489, 1129

\bibitem[\protect\citeauthoryear{Binney et al.}{2000}]{Binney00} 
    Binney, J., Dehnen, W., \& Bertelli, G. 2000, MNRAS, 318, 658

\bibitem[\protect\citeauthoryear{Blitz et al.}{1984}]{Blitz84} 
   Blitz, L., Magnani, L. \& Mundy, L., 1984, ApJ, 282, 9

\bibitem[\protect\citeauthoryear{Blitz et al.}{1999}]{Blitz99} 
   Blitz L, Spergel D.N., Teuben P.J., Hartmann D. \& Burton W.B. 1999, ApJ, 514, 818

\bibitem[\protect\citeauthoryear{Bonatto \& Bica}{2007b}]{Bonatto07}
   Bonatto C. \& Bica E. 2007b, MNRAS, 377, 1301

\bibitem[\protect\citeauthoryear{Bonatto \& Bica}{2008}]{StrucPar}
   Bonatto C. \& Bica E. 2008, A\&A, 477, 829

\bibitem[\protect\citeauthoryear{Bonatto \& Bica}{2008}]{Bonatto08}
   Bonatto C. \& Bica E. 2008, A\&A, 485, 81

\bibitem[\protect\citeauthoryear{Bonatto \& Bica}{2009}]{Bonatto09}
   Bonatto C. \& Bica E. 2009, MNRAS, 397, 1915

\bibitem[\protect\citeauthoryear{Bonatto \& Bica}{2010}]{Bonatto10} 
   Bonatto, C. \& Bica, E. 2010, A\&A, 516, 81

\bibitem[\protect\citeauthoryear{Bonatto \& Bica}{2011b}]{Bonatto11}
   Bonatto C. \& Bica E. 2011b, A\&A, 530, 32

\bibitem[\protect\citeauthoryear{Bregman}{1980}]{Bregman80}
   Bregman, J.~N., 1980, ApJ, 236, 577

\bibitem[\protect\citeauthoryear{Bressan et al.}{2012}]{Bressan12} 
   Bressan, A., Marigo, P., Girardi, L., Salasnich, B., Dal Cero, C., Rubele, S., \& Nanni, A. 2012, MNRAS, 427, 127

\bibitem[\protect\citeauthoryear{Camargo et al.}{2009}]{Camargo09} 
   Camargo, D., Bonatto, C. \& Bica, E. 2009, A\&A, 508, 211

\bibitem[\protect\citeauthoryear{Camargo et al.}{2010}]{Camargo10} 
   Camargo, D., Bonatto, C. \& Bica, E. 2010, A\&A, 521, 42

\bibitem[\protect\citeauthoryear{Camargo et al.}{2011}]{Camargo11} 
   Camargo, D., Bonatto, C. \& Bica, E. 2011, MNRAS, 416, 1522

\bibitem[\protect\citeauthoryear{Camargo et al.}{2012}]{Camargo12} 
   Camargo, D., Bonatto, C. \& Bica, E. 2012, MNRAS, 423, 1940

\bibitem[\protect\citeauthoryear{Camargo et al.}{2013}]{Camargo13} 
   Camargo, D., Bica, E. \& Bonatto, C. 2013, MNRAS, 432, 3349

\bibitem[\protect\citeauthoryear{Camargo et al.}{2015b}]{Camargo15a} 
   Camargo, D., Bica, E. \& Bonatto, C. 2015b, NewA, 34, 84

\bibitem[\protect\citeauthoryear{Camargo et al.}{2015a}]{Camargo15b} 
   Camargo, D., Bica, E., Bonatto, C. \& Salerno, G. 2015a, MNRAS, 448, 1930

\bibitem[\protect\citeauthoryear{Camargo et al.}{2015c}]{Camargo15c} 
   Camargo, D., Bonatto, C. \& Bica, E. 2015c, MNRAS, 450, 4150

\bibitem[\protect\citeauthoryear{Camargo et al.}{2016}]{Camargo16} 
   Camargo, D., Bica, E. \& Bonatto, C. 2016, MNRAS, 455, 3126

\bibitem[\protect\citeauthoryear{Chiappini et al.}{2001}]{Chiappini01} 
   Chiappini, C., Matteucci, F., \& Romano, D. 2001, ApJ, 554, 1044

\bibitem[\protect\citeauthoryear{Christodoulou et al.}{1997}]{Christodoulou97}
   Christodoulou D. M., Tohline J. E., \& Keenan F. P., 1997, ApJ, 486, 810

\bibitem[\protect\citeauthoryear{Cresci et al.}{2010}]{Cresci10} 
   Cresci, G., Mannucci, F., Maiolino, R., Marconi, A., Gnerucci, A. \& Magrini, L., 2010, Natur, 467, 811

\bibitem[\protect\citeauthoryear{Dickey \& Lockman}{1990}]{Dickey90}
   Dickey J. M., Lockman F. J., 1990, ARA\&A, 28, 215

\bibitem[\protect\citeauthoryear{Drlica-Wagner et al.}{2015}]{Drlica15} 
   Drlica-Wagner, A., Bechtol, K., Rykoff, E.~S., Luque, E., Queiroz, A., Mao, Y.-Y., Wechsler, R.~H., Simon, J.~D. et al. 2015, ApJ, 813, 109

\bibitem[\protect\citeauthoryear{Fern{\'a}ndez et al.}{2012}]{Fernandez12} 
   Fern{\'a}ndez, X., Joung, M.~R. \& Putman, M.~E. 2012, ApJ, 749, 181

\bibitem[\protect\citeauthoryear{Fraternali et al.}{2001}]{Fraternali01} 
    Fraternali, F., Oosterloo, T., Sancisi, R., \& van Moorsel, G. 2001, ApJ, 562, L47

\bibitem[\protect\citeauthoryear{Fraternali \& Tomassetti}{2012}]{Fraternali12} 
    Fraternali, F., \& Tomassetti, M. 2012, MNRAS, 426, 2166 

\bibitem[\protect\citeauthoryear{Fraternali}{2014}]{Fraternali14} 
    Fraternali F., 2014, in IAU Symposium, edited by S. Feltzing, G. Zhao, N. A. Walton, P. Whitelock, vol. 298 of IAU Symposium, 228–239

\bibitem[\protect\citeauthoryear{Hammer et al.}{2015}]{Hammer15} 
   Hammer, F., Yang, Y.~B., Flores, H., Puech, M. \& Fouquet, S. 2015, ApJ, 813, 110

\bibitem[\protect\citeauthoryear{Heald et al.}{2007}]{Heald07} 
    Heald, G. H., Rand, R. J., Benjamin, R. A., \& Bershady, M. A. 2007, ApJ, 663, 933

\bibitem[\protect\citeauthoryear{Heiles et al.}{1988}]{Heiles88} 
   Heiles, C., Reach,W. T. \& Koo, B. C. 1988, ApJ, 332, 313

\bibitem[\protect\citeauthoryear{Heitsch \& Putman}{2009}]{Heitsch09}
   Heitsch, F. \& Putman, M.~E., 2009, ApJ, 698, 1485

\bibitem[\protect\citeauthoryear{Hernandez et al.}{2013}]{Hernandez13} 
   Hernandez A. K., Wakker B. P., Benjamin R. A., French D., Kerp J., Lockman F. J., O’Toole S. \& Winkel B. 2013, ApJ, 777, 19

\bibitem[\protect\citeauthoryear{Huxor \& Grebel}{2015}]{Huxor15}
   Huxor, A. P., \& Grebel, E. K. 2015, MNRAS, 453, 2653

\bibitem[\protect\citeauthoryear{Joung et al.}{2012}]{Joung12} 
   Joung, M.~R., Bryan, G.~L., Putman, M.~E. 2012, ApJ, 745, 148

\bibitem[\protect\citeauthoryear{Kaufmann et al.}{2006}]{Kaufmann06} 
   Kaufmann, T., Mayer, L., Wadsley, J., Stadel, J. \& Moore, B., 2006, MNRAS, 370, 1612

\bibitem[\protect\citeauthoryear{Kaufmann et al.}{2010}]{Kaufmann10}
   Kaufmann, G., Li, C. \& Heckman, T. 2010, MNRAS, 409, 491

\bibitem[\protect\citeauthoryear{Ker{\'e}s et al.}{2005}]{Keres05}
   Ker{\'e}s D., Katz N., Weinberg D. H. \& Dav{\ ́e} R., 2005, MNRAS, 363, 2

\bibitem[\protect\citeauthoryear{King}{1962}]{King62} 
   King, I. 1962, AJ, 67, 471

\bibitem[\protect\citeauthoryear{Koposov et al.}{2015}]{Koposov15} 
   Koposov, S.~E., Belokurov, V., Torrealba, G. \& Evans, N.~W. 2015, ApJ, 805, 130

\bibitem[\protect\citeauthoryear{Kuntz \& Danly}{1996}]{Kuntz96}
   Kuntz, K. D., \& Danly, L. 1996, ApJ, 457, 703


\bibitem[\protect\citeauthoryear{Lada \& Lada}{2003}]{Lada03} 
   Lada, C.J. \& Lada, E.A. 2003, ARA\&A, 41, 57

\bibitem[\protect\citeauthoryear{Lee \& Chen}{2009}]{Lee09}
   Lee, H.-T. \& Chen, W.~P. 2009, ApJ, 694, 1423

\bibitem[\protect\citeauthoryear{Luque et al.}{2016}]{Luque16}
   Luque, E., Queiroz, A., Santiago, B., Pieres, A., Balbinot, E., Bechtol, K., Drlica-Wagner, A., Neto, A.~F., et al. 2016, MNRAS, 458, 603

\bibitem[\protect\citeauthoryear{Maller \& Bullock}{2004}]{Maller04}
   Maller A. H. \& Bullock J. S., 2004, MNRAS, 355, 694

\bibitem[\protect\citeauthoryear{Magnani et al.}{1996}]{Magnani96}
   Magnani, L.; Hartmann, D.; Speck, B. G. 1996, ApJS, 106, 447

\bibitem[\protect\citeauthoryear{Majaess}{2013}]{Majaess13} 
   Majaess, D. 2013 Ap\&SS, 344, 175

\bibitem[\protect\citeauthoryear{Martin et al.}{2015}]{Martin15} 
   Martin, P.~G., Blagrave, K.~P.~M., Lockman, F.~J., Pinheiro Gon{\c c}alves, D., Boothroyd, A.~I., Joncas, G. Miville-Desch{\^e}nes, M.-A. \& Stephan, G. 2015 ApJ 809 153

\bibitem[\protect\citeauthoryear{Melioli et al.}{2008}]{Melioli08}
   Melioli, C., Brighenti, F., D'Ercole, A. \& de Gouveia Dal Pino, E.~M., 2008, MNRAS, 388, 573

\bibitem[\protect\citeauthoryear{Melioli et al.}{2009}]{Melioli09}
   Melioli, C., Brighenti, F., D'Ercole, A. \& de Gouveia Dal Pino, E.~M., 2009, MNRAS, 399, 1089

\bibitem[\protect\citeauthoryear{Muller et al.}{1963}]{Muller63}
   Muller, C. A., Oort, J. H., \& Raimond , E. 1963, C. R. Acad. Sci. Paris, 257, 1661

\bibitem[\protect\citeauthoryear{Nichols \& Bland-Hawthorn}{2009}]{Nichols09}
   Nichols, M. \& Bland-Hawthorn, J. 2009, ApJ, 707, 1642

\bibitem[\protect\citeauthoryear{Oort}{1966}]{Oort66}
   Oort, J. H. 1966, Bull. Astron. Inst. Neth., 18, 421

\bibitem[\protect\citeauthoryear{Oosterloo et al.}{2007}]{Oosterloo07}
   Oosterloo, T., Fraternali, F., \& Sancisi, R. 2007, AJ, 134, 1019

\bibitem[\protect\citeauthoryear{Perryman et al.}{2001}]{Perryman01}
   Perryman, M. A. C., de Boer, K. S., Gilmore, G., et al. 2001, A\&A, 369, 339

\bibitem[\protect\citeauthoryear{Pl{\"o}ckinger \& Hensler}{2012}]{Plockinger12}
   Pl{\"o}ckinger, S. \& Hensler, G. 2012, A\&A, 547, 43

\bibitem[\protect\citeauthoryear{Pidopryhora et al.}{2007}]{Pidopryhora07}
   Pidopryhora, Y., Lockman, F.~J. \& Shields, J.~C., 2007, ApJ, 656, 928

\bibitem[\protect\citeauthoryear{Putman et al.}{2012}]{Putman12}
   Putman, M.E., Peek, J.E.G., \& Joung, M.R. 2012, ARA\&A, 50, 491

\bibitem[\protect\citeauthoryear{Putman et al.}{2004}]{Putman04}
   Putman, M. E., Thom, C., Gibson, B. K., \& Staveley-Smith, L. 2004, ApJ, 603, 77

\bibitem[\protect\citeauthoryear{Quilis \& Moore}{2001}]{Quilis01}
   Quilis, V. \& Moore, B. 2001, ApJ, 555, 95

\bibitem[\protect\citeauthoryear{Sancisi et al.}{2008}]{Sancisi08}
   Sancisi, R., Fraternali, F., Oosterloo, T. \& van der Hulst, T. 2008, A\&AR, 15, 189

\bibitem[\protect\citeauthoryear{Schlafly et al.}{2015}]{Schlafly15}
   Schlafly, E.~F., Green, G., Finkbeiner, D.~P., Rix, H.-W., Burgett, W.~S., Chambers, K.~C., Draper, P.~W., Kaiser, N. et al. 2015, ApJ, 799, 1165

\bibitem[\protect\citeauthoryear{Shapiro \& Field}{1976}]{Shapiro76}
   Shapiro, P.~R. \& Field, G.~B., 1976, ApJ, 205, 762

\bibitem[\protect\citeauthoryear{Smith}{1963}]{Smith63}
   Smith G. P., 1963, Bull. Astron. Inst. Neth., 17, 203

\bibitem[\protect\citeauthoryear{Spitoni et al.}{2009}]{Spitoni09}
   Spitoni, E., Matteucci, F., Recchi, S., Cescutti, G. \& Pipino, A., 2009 A\&A 504 87

\bibitem[\protect\citeauthoryear{Tepper-Garcia et al.}{2015}]{Tepper15}
   Tepper-Garcia, T., Bland-Hawthorn, J. \& Sutherland, R.~S. 2015, ApJ, 813, 94

\bibitem[\protect\citeauthoryear{Thilker et al.}{2004}]{Thilker04}
   Thilker, D.~A., Braun, R., Walterbos, R.~A.~M., Corbelli, E., Lockman, F.~J., Murphy, E. \& Maddalena, R. 2004, ApJ, 601, 39

\bibitem[\protect\citeauthoryear{Tutukov}{1978}]{Tutukov78} 
   Tutukov A.V. 1978, A\&A, 70, 57

\bibitem[\protect\citeauthoryear{Wolfe et al.}{2015}]{Wolfe15}
   Wolfe, S.A., Lockman, F.J. \& Pisano, D.J. 2016, ApJ, 816, 81 

\bibitem[\protect\citeauthoryear{Wright et al.}{2010}]{Wright10}
   Wright, E. L., Eisenhardt, P. R. M., Mainzer, A. K. et al. 2010, AJ, 140, 1868

\end{thebibliography}
\end{document}